# Magnetic resonance imaging of mean cell size in human breast tumors

Junzhong Xu[1,2,3,4*], Xiaoyu Jiang[1,2], Hua Li[1,2], Lori R. Arlinghaus[1,2], Eliot T. McKinley[5], Sean P. Devan[1], Benjamin M. Hardy[1,4], Jingping Xie[1,2], Hakmook Kang[6], Anuradha B. Chakravarthy[7], John C. Gore[1,2,3,4]

[1] Institute of Imaging Science, Vanderbilt University Medical Center, Nashville, TN 37232, USA

[2] Department of Radiology and Radiological Sciences, Vanderbilt University Medical Center, Nashville, TN 37232, USA

[3] Department of Biomedical Engineering, Vanderbilt University, Nashville, TN 37232, USA

[4] Department of Physics and Astronomy, Vanderbilt University, Nashville, TN 37232, USA

[5] Department of Medicine, Vanderbilt University Medical Center, Nashville, TN 37232, USA.

[6] Department of Biostatistics, Vanderbilt University Medical Center, Nashville, TN 37232, USA.

[7] Department of Radiation Oncology, Vanderbilt University Medical Center, Nashville, TN 37232, USA.

[*] **Corresponding author**: Address: Vanderbilt University, Institute of Imaging Science, 1161 21st Avenue South, AAA 3113 MCN, Nashville, TN 37232-2310, United States. Fax: +1 615 322 0734. E-mail: junzhong.xu@vanderbilt.edu (Junzhong Xu) Twitter: @JunzhongXu.

**Running Title**: MRI of human breast tumor cell size

## Abstract

**Purpose**: Cell size is a fundamental characteristic of all tissues, and changes in cell size in cancer reflect tumor status and response to treatments, such as apoptosis and cell cycle arrest. Unfortunately, cell size can only be obtained by pathologic evaluation of the tumor in the current standard of care. Previous imaging approaches can be implemented on only animal MRI scanners or require relatively long acquisition times that are undesirable for clinical imaging. There is a need to develop cell size imaging for clinics.

**Experimental Design**: We propose a new method, IMPULSED (Imaging Microstructural Parameters Using Limited Spectrally Edited Diffusion) that can characterize mean cell sizes in solid tumors. We report the use of combined sequences with different gradient waveforms on human MRI and analytical equations that link DWI signals of real gradient waveforms and specific microstructural parameters such as cell size. We also describe comprehensive validations using computer simulations, cell experiments *in vitro*, and animal experiments *in vivo* and finally demonstrate applications in pre-operative breast cancer patients.

**Results**: With fast acquisitions (~ 7 mins), IMPULSED can provide high-resolution (1.3 mm in-plane) mapping of mean cell size of human tumors in vivo on currently-available 3T MRI scanners. All validations suggest IMPULSED provide accurate and reliable measurements of mean cell size.

**Conclusion**: The proposed IMPULSED method can assess cell size variations in the tumor of breast cancer patients, which may have the potential to assess early response to neoadjuvant therapy.

**Introduction**

Cell size is a basic feature of living cells that plays an important role from the molecular to the organismal level, including cellular metabolism, (1) proliferation, (2) and tissue growth (3). Cell size may vary significantly during disease progression or after therapy. For example, cells swell significantly after acute stroke (4,5) and cell shrinkage is a hallmark of apoptotic cell death (6). For cancer diagnosis and prognosis, cell size is of interest because it varies during mitosis and before death, and thus may provide a unique means to evaluate tumor progression and response to treatments. Measurements of cell sizes are reportedly capable of differentiating cancer types (7) and monitoring tumor early therapeutic response by detecting treatment-induced apoptosis (8,9) or mitotic-arrest (10). Therefore, quantitative microstructural measurements such as cell size may provide specific means to probe the status of cancerous tissues and would be of potential value in preclinical and clinical applications. Currently, such microstructural information is obtained with conventional clinical care only via invasive biopsies, which are limited not only by the potential to miss important changes due to tumor heterogeneity and the small sample size of each specimen, but also may introduce various clinical complications, including pain, hemorrhage, infection, and even death (11). Therefore, a non-invasive imaging technique capable of characterizing tissue microstructural information would be of great interest to clinicians.

Diffusion-weighted magnetic resonance imaging (DWI) is an exogenous-agent-free non-invasive imaging technique that provides a unique capability to probe biological tissue microstructure by evaluating the degree of restriction and hindrance to the free motion of randomly diffusing water molecules. Values of the apparent diffusion coefficient (ADC), a metric obtained using DWI, have been found to be sensitive to cell density and hence are widely used to evaluate cellularity changes after anti-cancer treatment (12-15). However, ADC represents an overall diffusion property of water molecules inside each image voxel, and ADC values are influenced by several tissue parameters simultaneously, including but not limited to cell size (16), cell membrane permeability (17), intra- and extracellular diffusion coefficients (18), and intracellular volume fraction (19). As a result, ADC and tumor cellularity are not always strongly correlated (20-22) so that ADC does not reliably provide specific information on cell size and density. Recently, numerous attempts have been made to enhance the specificity of DWI measurements, such as the DDR (23), VERDICT (24-26), qTDS (27,28), and POMACE approaches (29). Some of these e.g. qTDS, exploit the dependence of ADC on the time scale (the diffusion time) over which diffusion affects the measured signals, which in principle enables the derivation of the spatial scales of restrictions to free displacements (27,28). However, previous methods were either implemented on animal MRI scanners only and used much stronger diffusion gradient amplitudes than are available on clinical MRI scanners, or the total acquisition times were long that is undesirable for clinical practice. Therefore, there is still a need to develop a fast (< 10 mins) and quantitative DWI method that is capable of measuring cell size on clinical MRI machines.

Here, we introduce a modified qTDS approach termed IMPULSED for clinical MR imaging. Similar to our previously developed approach that used apodised cosine oscillating diffusion gradients (27,28), we propose IMPULSED acquisitions to incorporate cosine-modulated oscillating trapezoidal diffusion gradients (30,31) that are readily capable of running on clinical MRI machines as shown in Figure 1. An IMPULSED protocol incorporates a set of diffusion weighted imaging sequences, each of which uses a different diffusion time, with a diffusion time range that makes the DWI images highly sensitive to variations in cell size in human tissues, and from which 3D maps of cell size and density can be derived. For human breast tumors, all the necessary data can be acquired in scan times of < 7 mins. In the current work, we demonstrate a practical imaging protocol that combines different DWI sequences along with new analytical equations that link the DWI signals using the real gradient waveforms to specific microstructural parameters such as cell size. We also provide validations of the IMPULSED method using computer simulations and experimental measurements of cells *in vitro* and animal models *in vivo*, along with practical demonstrations of clinical applications in breast cancer patients.

**Materials and Methods**

**Pulse sequence**

Figure 1 shows the pulse sequences used to acquire DWI data for the IMPULSED method.

In addition to a conventional pulsed gradient spin echo (PGSE) sequence which measures ADC over longer diffusion times, IMPULSED also uses cosine-modulated, trapezoidal, oscillating gradients in spin echo sequences (OGSE) to measure ADC over different, shorter diffusion times. The combination of longer and shorter diffusion times ensures that ADC values from each sequence will differ, and these differences then reflect the length scales of major restrictions to free diffusion, which in tumors correspond to cell sizes (32). This combination enables detection of a broader range of length scales, providing more comprehensive information on tissue microstructure than single measurements of ADC (27,28). For all diffusion sequences, $G$ is the gradient strength, $\delta$ is the duration of each diffusion gradient, $\Delta$ is the separation of two gradients, $t_r$ is the gradient rise time (= 0.9 ms on Philips Achieva 3T MRI scanner), $t_p$ is the duration time of the first gradient plateau, and $t_3 = t_p + t_r/2$ for OGSE sequences. $N$ is the number of cycles in each diffusion gradient in the OGSE sequence. $b = \gamma^2 G^2 \left[ (t_r + t_p)^2 (\Delta - (t_r + t_p)/3) + t_r^3/30 - (t_r + t_p) t_r^2 / 6 \right]$ for the PGSE sequence and $b = \gamma^2 G^2 \left[ 91 N t_r^3 / 15 + 8 N t_p^3 / 3 + t_r^3 / 30 + 12 N t_r t_p^2 + 46 N t_r^2 t_p / 3 \right]$ for the OGSE sequences (30,31) shown in Figure 1.

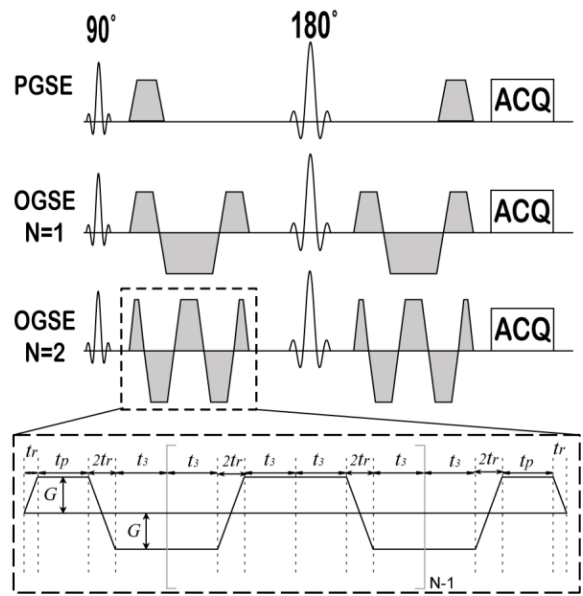

Figure 1. Diagram of the pulse sequences used in the IMPULSED method. In addition to conventional PGSE acquisitions, OGSE acquisitions with two frequencies ($N$ = 1 and 2) are used.

**Theory**

Quantitative information on cell size and density are obtained by fitting a simple model of tissue water to the DWI experimental data. Like previous reports (24,27,29,33), DWI signals in IMPULSED are modeled as the sum of signals arising from two compartments, i.e., intracellular and extracellular spaces. The measured DWI signal $S$ is

$$S = v_{in} \cdot S_{in} + (1 - v_{in}) \cdot S_{ex} \quad [1]$$

where $S_{in}$, and $S_{ex}$ are the signal magnitudes per volume from the intracellular and extracellular spaces, respectively, and $v_{in}$ is intracellular water fraction. Note that water exchange between intra- and extracellular spaces is ignored but does not affect the estimation of mean cancer cell size although it may bias the estimation of cell density (34).

***Modeling intracellular diffusion*** (the 1[st] term in Eq.[1]): Water molecules are restricted inside cells because cell membranes have only finite permeabilities. For simplification, cancer cells are usually modeled as impermeable spheres and analytical expressions similar to previous reports (33) may be derived to link DWI signals and underlying microstructural parameters such as cell size. The derived analytical equations for cosine-modulated, trapezoidal-shaped OGSE sequences, and a corresponding validation of their accuracy using computer simulations, can be found in the Supplemental Materials.

***Modeling extracellular diffusion*** (the 2[nd] term in Eq.[1]). It is challenging to obtain an explicit analytical form to describe extracellular diffusion. When the diffusion time range is limited, previous studies have suggested the extracellular diffusion coefficient shows an approximately linear dependence on the gradient frequency (proportional to the inverse of diffusion time) (23,35). However, due to hardware limitations, the frequency range available on clinical MRI scanners is narrow so that the dependency of extracellular diffusion on frequency is minor. All our investigations using simulations *in silico*, cell lines *in vitro*, and animals *in vivo* have suggested the extracellular diffusion coefficient obtained using the IMPULSED method is largely insensitive to diffusion times so that the extracellular diffusion coefficient is modeled as a constant $D_{ex}$.

***Diffusion anisotropy.*** Except for some *ex vivo* investigations of fixed breast tissues with strong gradients (36), a previous *in vivo* study involving 81 patients reported that water diffusion is nearly isotropic in various human breast tumors (37). Therefore, the DWI signals for IMPULSED acquisitions are obtained by averaging three acquisitions with diffusion gradients along three orthogonal oblique directions ([$g_x$, $g_y$, $g_z$] = [1, 1, -0.5], [1, -0.5, 1], and [-0.5, 1, 1]), corresponding to the trace of a diffusion tensor.

*IMPULSED outcomes*: Table 1 lists tissue parameters that can be determined by analyses of DWI data. The estimates of $d$ are actually volume-weighted mean cell sizes (38) which, for a population of cells, is given by $d = \sum_n d_n^4 / \sum_n d_n^3$, where $d_n$ is the cell size of the $n^{th}$ cell. The cell density $\rho$ can be calculated from $d$ and $v_{in}$ i.e., $\rho = \frac{6v_{in}}{\pi d^3}$. Note that, unlike our pre-clinical animal studies (27,28), $D_{in}$ is fixed as 1.58 µm²/ms (10) in order to better stabilize the fittings of IMPULSED. Our simulations suggest that difference choices of $D_{in}$ in the data fitting has little influences on the fitted IMPULSED metrics (see Figure S5 in the Supplemental Materials).

Table 1 Summary of IMPULSED derived parameters and corresponding biophysical features.

| IMPULSED parameters | Biophysical features |
|---|---|
| $d$ | Volume-weighted mean cell size |
| $v_{in}$ | Intracellular volume fraction |
| $D_{ex}$ | Extracellular diffusion coefficient |

**Experimental diffusion parameters in IMPULSED**

It is desirable to use higher gradient strengths and slew rates in diffusion measurements to achieve greater diffusion weighting (b values) and a range of diffusion times appropriate for each sample of interest (39). However, due to human physiological thresholds and hardware limitations, the gradient strength and slew rate are limited on clinical MRI scanners, which makes it challenging to implement some quantitative DWI methods. To ensure IMPULSED can be translated clinically, a maximum gradient strength of 60 mT/m on any single axis and a slew rate

< 100 mT/m was assumed, and these limits were imposed on our simulations and experiments including cells, animals, and humans. Table 2 shows diffusion parameters for IMPULSED measurements that are readily available on human 3T MRI systems and were selected for practical implementation.

Table 2 Summary of diffusion parameters used in IMPULSED measurements which are readily available for human 3T MRI system.

|  | δ /Δ [ms] | $N$ | $f$ [Hz] | $b$ [s/mm$^2$] |
|---|---|---|---|---|
| **PGSE** | 12/74 | N/A | N/A | 0,250,500,750,1000, 1400, 1800 |
| **OGSE** | 40.9/51.4 | 1 | 25 | 0,250,500,750,1000 |
|  |  | 2 | 50 | 0,100,200,300 |

**Validation using simulations *in silico***

Computer simulations were performed to evaluate both the accuracy and precision of IMPULSED derived parameters obtainable for signal to noise ratios (SNRs) practically available on clinical 3T MRI scanners. A finite difference method was used to simulate DWI signals obtained using PGSE and OGSE sequences as reported previously (40). Tumors were modeled as tightly-packed spherical cells on a face-centered-cubic lattice (41) with $f_{in}$ = 61.8%, $D_{in}$ = 1.58 μm$^2$/ms (10), $D_{ex}$ = 2 μm$^2$/ms, and homogeneous relaxation times everywhere for simplicity. Eight different values of cell diameter $d$ evenly distributed from 6 to 20 μm were evaluated, covering the cell sizes typical of lymphocytes to cancer cells. After noise-free DWI signals were calculated for each cell diameter, Rician noise equivalent to achieve an SNR = 20 was added to mimic realistic DWI signals, and then the noisy signals were used for data fitting. This process was repeated 100 times so that the accuracy of each IMPULSED derived metric could be evaluated using the mean fitted values, and the precision evaluated using the corresponding standard deviations.

**Validation using cell lines *in vitro***

***Cell preparation***. Three types of breast cancer cell lines, MDA-MB-231, MCF7, and MDA-

MB-453, were purchased from American Type Culture Collection (Manassas, Virginia, USA). Cells were cultured in RPMI Medium 1640 supplemented with 10% FBS, 50 Units/ml penicillin, 50ug/ml streptomycin and 5ug/ml recombinant insulin (Invitrogen, CA) in a humidified incubator maintained with 5% $CO_2$ at 37°C. Cells were cultured in 150 mm dishes to full confluence, then harvested by trypsinization, washed and resuspended with PBS. A small portion of cells was aliquoted and fixed with 70% cold ethanol for flow cytometry analysis, and the majority of cells were fixed with 4% paraformaldehyde in PBS for over 2 hours for MR experiments and microscopic size analysis.

In addition to cancer cell lines, the Jurkat acute T cell leukemia cell line and lymphocytes were used to mimic smaller cells. Lymphocytes were extracted from human peripheral blood by using the Ficoll method (42); briefly, blood was diluted with an equal volume of PBS, and carefully added on top of an equal volume of Ficoll-Paque in a centrifuge tube. After centrifugation at 800g for 20 minutes, the cells in the interface layers were collected, the residual red cells were removed by hypotonic lysis and washing, and the final lymphocytes were pelleted and re-suspended with PBS.

For MR experiments, cells were washed with PBS after fixation, about $3 \times 10^7$ cultured cells (or $1 \times 10^9$ lymphocytes) were centrifuged at 2000g for 2 minutes in a 0.65ml of Eppendorf tube to obtain a tight cell pellet. The supernatant was carefully removed for MRI measurements. A small aliquot of cells from each sample was spotted on a glass slide and covered by a coverslip. Digital images of the cells were recorded at both 20X and 40X amplification. A stage micrometer was used for size calibration.

***MRI experiments***.   All MRI measurements of cells *in vitro* were performed on a Varian 4.7T MRI spectrometer similar to the approach described previously (34). A 2-mm thick slice crossing the center of each cell pellet was imaged with a field-of-view 16×16mm and a matrix size 64×64, yielding a spatial resolution of 250μm. All diffusion sequence parameters were the same as in Table 2.

***Light microscopy***.   Bright field images were captured using a Zeiss Axio Observer microscope. Two representative fields of view were chosen for each slide. Each field of view was imaged at 40X magnification with the focus set slightly above, below, and equal to the optimal focal plane. Differences between two out-of-focus images resulted in enhanced contrast of the cell boundaries (43). The area of each cell was then calculated from these microscopic images using an auto-segmentation program written in Matlab, and these measurements were converted to a diameter assuming each cell is a sphere. The detailed procedures are provided in the Supplemental Materials and representative images are shown in Figure S6.

**Validation using animals *in vivo***

*Animal preparation*. All procedures were approved by the Vanderbilt University Institutional Animal Care and Usage Committee. Three MDA-MB-231 and two MCF7 xenografts were generated following subcutaneous injection of $1-2 \times 10^6$ cells in female athymic nude mice (Harlan Laboratories, Inc., Indianapolis, IN). When each tumor reached a size of 200–300 mm$^3$, MR imaging was performed as described below, and each mouse was euthanized for histology immediately afterward.

*MRI experiments*. All MR images of animals were acquired on a 4.7 T Varian horizontal small animal scanner using a Litz38 volume coil for both transmission and reception. A single-shot echo-planar imaging (EPI) diffusion sequence with fat suppression was used for all diffusion measurements to minimize motion artefacts. Axial slices with 1 mm thickness were acquired to cover the entire tumors, with in-plane matrix size 128 × 64 and field-of-view =40 × 20 mm yielding an in-plane resolution of 312.5 × 312.5 µm. All diffusion sequence parameters are outlined in Table 2.

*Histology*. Animals were euthanized immediately following MRI scans. Tumors were collected, cut into 2mm thick pieces and fixed in 10% formalin for 24 hours. Tissues were transferred to 70% ethanol prior to paraffin embedding. Tumors were sectioned at 8 µm thickness and stained with hematoxylin and eosin (H&E) for cellularity, and Na$^+$/K$^+$-ATPase (ab76020, Abcam). For Na$^+$/K$^+$-ATPase staining, slides were de-paraffinized, rehydrated, and antigen retrieved using 6.1 pH citrate buffer in a pressure cooker at 105°C for 20 minutes, followed by a

10-minute cool down. Slides were treated with a 3% hydrogen peroxide solution for 15 minutes and blocked in phosphate buffered saline/3% bovine serum albumin/ 10% donkey serum for 30 minutes prior to antibody staining. The primary antibody was incubated at 4°C overnight followed by secondary detection with a Cy5-labelled rabbit-specific antibody. Nuclei were stained with 40,6-diamidine-2-phenylidole-dihydrochloride (DAPI). Digital fluorescent images were collected on an Olympus IX-81 microscope with a magnification of 20. These images were analyzed using CellProfiler<sup>TM</sup> ([http://www.cellprofiler.org/](http://www.cellprofiler.org/)) to obtain quantitative information on cell sizes. The complete processing pipeline included illumination correction, foreground objects identification, and splitting of clumped cells by watershedding. Finally, the cell segmentation results obtained by CellProfiler<sup>TM</sup> were manually corrected for over-segmented cell bodies.

**Applications in patients**

*Breast cancer patients*. The human imaging study was approved by the Institutional Review Board at Vanderbilt University Medical Center. Seven women (age 55.3±8.0) diagnosed with breast cancer with tumors 1cm or greater were recruited in this study. Written informed consents were received from participants prior to inclusion in the study. The detailed patient information and tumor types can be found in Table S1 in the Supplemental Materials.

*Human MR imaging*. IMPULSED imaging was performed using a Philips Achieva 3T scanner with a 16-channel breast coil. Acquisition sequence parameters were TR/TE=4500/103ms; FOV=192×192mm; reconstructed in-plane resolution = 1.3×1.3 mm; 10 or 20 slices; slice thickness=5 mm; single shot EPI; SENSE factor=3; fat suppression with SPAIR; and dynamic stabilization was used to minimize DWI signal drifts during scanning. Images were acquired with two opposite diffusion gradient directions for each axis and the geometric means were used as final images to mitigate the cross-terms between diffusion and background gradients (44). All diffusion sequence parameters were the same as in Table 2. The total scan time ≈ 7 mins. In addition, ADC measurements using PGSE acquisitions with $\Delta$ = 54 and 34 ms were performed to further investigate the ADC dependence on diffusion times.

**Data analyses**

For cell experiments *in vitro*, an ROI avoiding boundaries was manually drawn on each image and the total DWI signals from the ROI were used in the data fitting. The light microcopy images of cells were analyzed using a locally developed pipeline with details provided in the Supplemental Materials. For experiments in vivo, tumor ROIs were manually drawn based on PGSE diffusion weighted images with b = 1000 s/mm$^2$. The IMPULSED fitting was performed only inside ROIs. The histology images were analyzed using CellProfiler™ and the details are provided in the Supplemental Materials.

All data fittings were performed using Matlab (Mathworks, Natick, Massachusetts) to generate DWI parametric maps on a voxel-wise basis (38).   The fitting parameter ranges were limited by possible physiologically relevant values, i.e., 0.2 ≤ $d$ ≤ 25 μm (i.e., typical breast cancer cell size range. Note that the upper limit is determined by the root mean square displacement of free water diffusion at 37°C), 0 ≤ $v_{in}$ ≤ 1 (max volume fraction 100%), and 0 ≤ $D_{ex}$ ≤ 3.1 μm$^2$/ms (the free water diffusion coefficient is 3.07 μm$^2$/ms at 37°C). $D_{in}$ was fixed as 1.58 μm$^2$/ms (10) in the data analysis to stabilize fittings as shown in Figure S5. Note that fitting results are insensitive to the choices of $D_{in}$. The fittings were to maximize the log likelihood function with Rician noise, i.e., $L_R = \sum_{n=1}^{N} \left[ \log M_n - 2\log \sigma + \log I_0 \left( \frac{S_n M_n}{\sigma^2} \right) - \frac{S_n^2 + M_n^2}{2\sigma^2} \right]$ , where $S_n$ and $M_n$ are the model-predicted and measured signals of the $n^{th}$ measurements, respectively, and $\sigma^2$ is the noise variation. After data fitting, all diffusion images and IMPULSED derived parametric images were co-registered to corresponding high-resolution T1-weighted anatomical images using the FMRIB's Linear Image Registration Tool (FLIRT) (45) in the FSL toolbox (46).

The Spearman correlation was calculated to determine the relationship between fitted cell size and histology values. The Wilcoxon rank-sum test was used to evaluate differences of cell size values between those from fitted and those derived from histology.

## Results

### Computer simulations *in silico*

Figure 2 shows the simulated dependence of IMPULSED derived metrics on input values with a signal to noise ratio (SNR) of 20, typical of human MRI. In the range 6 – 20 µm, the fitted cell size $d$ shows a clear linear dependence on ground-truth values. A mixed linear model reports p<0.01. The fitted intracellular water fraction $v_{in}$ shows a good match to any ground-truth value with a Bonferroni adjusted p > 0.05. By contrast, the fitted extracellular diffusion coefficient $D_{ex}$ shows significant uncertainties, indicating the precision of fitted $D_{ex}$ is limited when SNR is low. Note that the fitted $D_{ex}$ is expected to be lower than the intrinsic extracellular diffusion coefficient due to restriction effects. If SNR increases to 50, the precisions of fitted $d$, $v_{in}$ and $D_{ex}$ can be dramatically improved (see Figure S3 in the Supplemental Materials). In conclusion, although limited in practice by the available gradient strength and slew rate, IMPULSED is capable of measuring mean cancer cell size reliably when 6 < $d$ < 20 µm with practical SNRs, but the estimation of extracellular diffusion coefficient is not reliable with low SNRs.

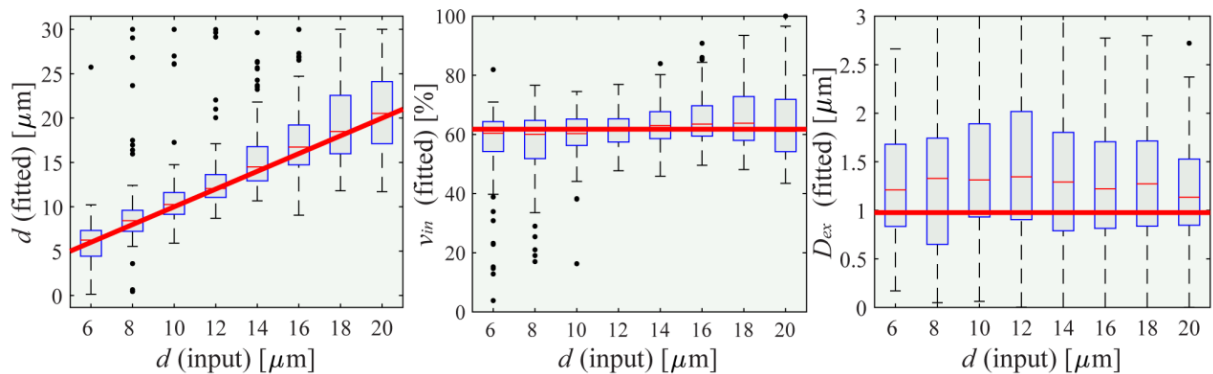

Figure 2 Simulated influence of noise with SNR = 20 on fitted $d$ (left), $v_{in}$ (middle), and $D_{ex}$ (rigth) using IMPULSED. For each real input $d$, the fittings were repeated 100 times each with different noise samples but with the same SNR level. The red solid lines represent the ground-truth values, boxes represent ranges between the 25th and 75th percentiles, and dots are outliers.

**Imaging cells *in vitro***

Figure S6 in the Supplemental Materials shows an example of the cell segmentation of light microscopy images. The cell size information obtained from microscopy was assumed to be the ground truth for validating the cell sizes fitted using IMPULSED. Figure 3 compares the mean cell sizes obtained from IMPULSED fitting and light microsocpy for three breast cancer cell lines, the Jarkat cell line, and lymphocytes. Over a broad range of cell sizes (11 – 18 µm) and cell types (breast cancer, leukemia, and lymphocytes), IMPULSED fitted $d$ values show a strong correlation (r = 0.92 and p < 0.001 provided by the Pearson correlation) with the values obtained from light microsocpy.

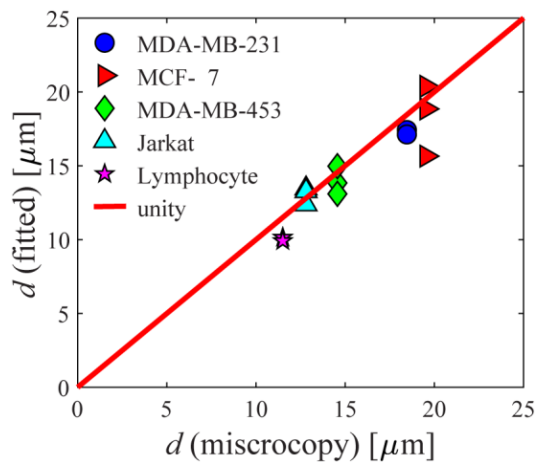

Figure 3 Comparison of mean cell sizes obtained from IMPULSED fitting and light microscopy.

**Imaging animals *in vivo***

Figure 4 shows a comparison of fitted and histology derived mean cell sizes for two types of breast cancer xenografts, MDA-MB-231 and MCF7. A Wilcoxon rank-sum test found no statistically significant differences (p>0.67) between values of both tumor types. Instead of pixel-wise analysis, the IMPULSED fitting of animal data *in vivo* was performed based on measurements from the entire tumors, while >15,700 cells were included in histological analyses for each tumor to minimize the effects of intra-tumor inhomogeneity. This validates the accuracy of IMPULSED derived mean cell size $d$ *in vivo*.

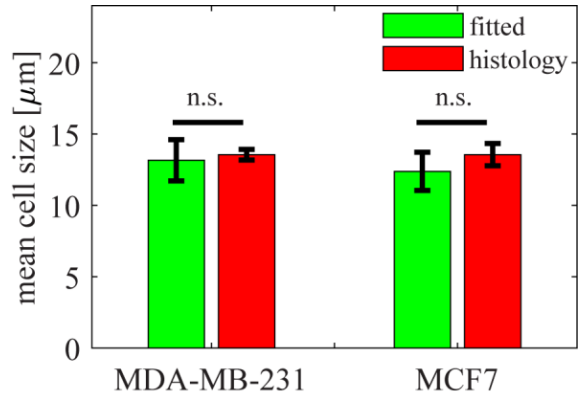

Figure 4 Comparison of fitted and histologically obtained mean cell sizes of two types of breast cancer xenografts. n.s.: no statistical significance.

**Imaging breast cancer patients *in vivo***

Figure 5 shows IMPULSED signals (symbols) from a tumor region of interest (ROI) and the IMPULSED fits (solid lines) of a representative human breast tumor. Note that for the same b values, signals acquired with OGSE with an effective diffusion time $t_{diff}$ of 10 ms decay significantly more than those obtained with PGSE with $t_{diff}$ = 70 ms. This increase in ADC at shorter diffusion times obtained provides the contrast that enables the possibility to measure cancer cell size and density. An example of ADC dependence on diffusion times can be found in Figure S8 in the Supplemental Materials. For the IMPULSED fitting, all b = 0 images were excluded from the fittings in order to minimize the influences of blood perfusion.

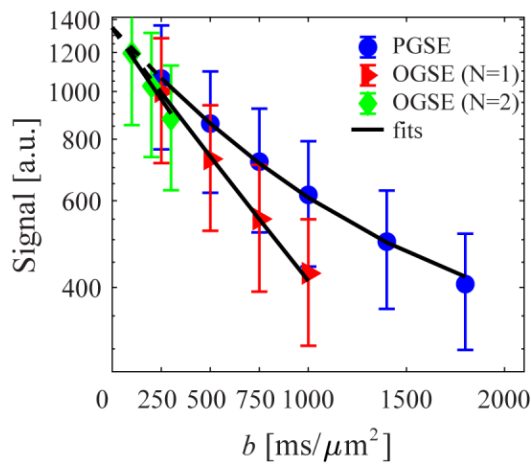

Figure 5 The ROI-based diffusion-weighted signal attenuations of a representative human breast

tumor. Markers are mean signals and the error bars represent standard deviations. The solid lines are fitted results using Eq.[1], and dashed lines with low b values indicate b=0 images were excluded from fittings to minimize perfusion effects.

Figure 6 shows representative IMPULSED-derived parametric maps of mean cancer cell size, intracellular volume fraction, and extracellular diffusion coefficient of a human breast tumor overlaid on a high-resolution anatomical image. The parametric maps are significantly inhomogeneous within the tumor. There are significantly lower intracellular volume fraction and increased extracellular diffusion coefficient at the center of the tumor, suggesting an acellular, necrotic core has developed. Note that the parametric map of $D_{ex}$ is more heterogeneous inside tumors which may be due to the lower fitting precision (see Figure 1).

For the tumor shown in Figure 6, the fitted overall average cell size $d$ = 14.88 ± 4.39 μm, $v_{in}$ = 38.87 ± 7.95%, and $D_{ex}$ = 1.81 ± 0.46 μm$^2$/ms, yielding a cell density = 2.86 ± 1.31 × 10$^8$ cells/cm$^3$, which is consistent with previous reports of ~10$^8$ cells/cm$^3$ for tumors of epithelial origin (such as breast tumors) (47). The histograms of all fitted IMPULSED metrics for seven breast cancer patients were summarized in Figure S9. All the fitted values of cell sizes, densities, and diffusion coefficients are within reasonable ranges.

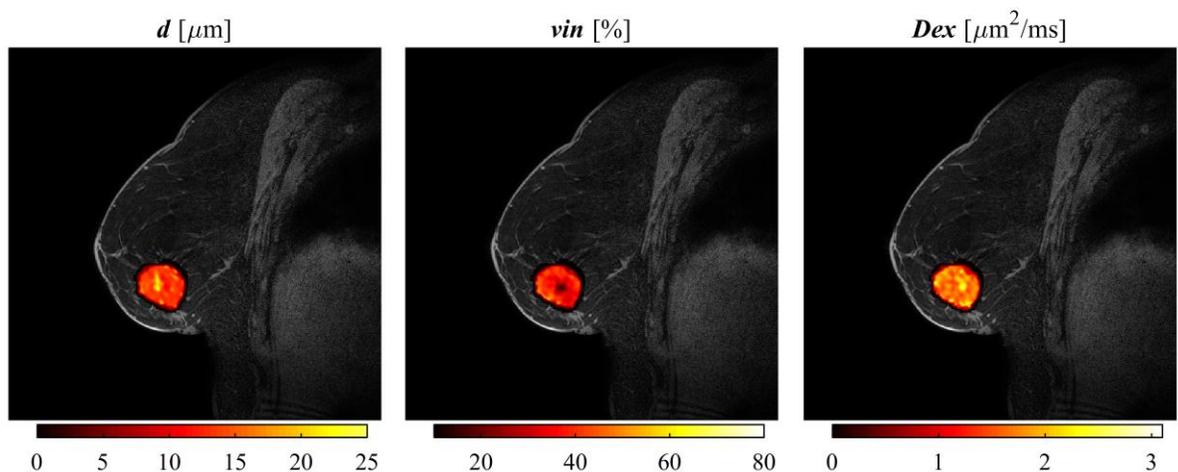

Figure 6 IMPULSED-derived maps of mean cell size $d$ (left), intracellular volume fraction $v_{in}$

(middle), and extracellular diffusion coefficient $D_{ex}$ (rigth) overlaid on a high-resolution fat-suppressed anatomical image of a breast cancer patient.

**Discussion**

We have developed and implemented a new, non-invasive imaging method termed IMPULSED to measure mean cancer cell size of tumors using clinical MRI scanners *in vivo*. Because cell size is one of the fundamental characteristics of cancer cells, the measurement of mean cell size provides opportunities to evaluate tumor status and monitor tumor therapeutic response. Compared with previous methods, IMPULSED has several advantages that make it particularly suitable for clinical translation: (1) the acquisition is fast (~ 7mins); (2) it is free of exogenous agents and radiation; and (3) can be implemented within the hardware performance specifications currently available on modern clinical 3T MRI scanners. Because MRI is already a part of standard-of-care management, IMPULSED has the potential to be added to clinical scan protocols to provide microstructural information that cannot be obtained by other methods easily.

Despite numerous attempts of using diffusion MRI to map microstructural information in tumors *in vivo*, successful applications in human tumors *in vivo* have so far been very limited. The VERDICT method is another diffusion MRI based technique to map mean cell size in human tumors, but to date has been exclusively used in human prostate cancers (24-26). Recent efforts at optimization have shortened its acquisition time from 35 minutes (25) to 12 minutes (26). The VERDICT approach is different from IMPULSED in that though VERDICT also employs multiple diffusion times, it uses conventional PGSE sequences with higher b values. Consequently, the range of diffusion times probed is narrower than for IMPULSED. Moreover, VERDICT uses multiple acquisitions with different echo times, which may introduce biases in fitting from relaxation time effects. IMPULSED incorporates cosine-modulated trapezoidal oscillating gradients to significantly increase the range of effective diffusion times in the short time range. Figure S8 in the Supplemental Materials shows the diffusion time dependence of ADC of a human breast tumor. Over the range $t_{diff}$ 30 – 70 ms achievable using PGSE sequences only (such as those used in VERDICT), the ADC does not change significantly. The most significant ADC change occurs between 30 ms and 5 ms, the range used in IMPULSED, so by measuring ADC using $t_{diff}$ over this range, the reliable derivation of cellular properties becomes practical.

Although IMPULSED detects the mean cell size within each image voxel, it is plausible to obtain the distribution of cell sizes inside each voxel as well. However, it is extremely challenging to fit cell size distributions without a *priori* knowledge of their nature (48). Most quantitative DWI studies to date have modeled cancer cells as spheres (10,27) or neural axons as cylinders (38) with uniform cell sizes to simplify the mathematical complexity yet preserve basic microstructural features. IMPULSED detects a mean cell size and density inside each imaging voxel, so the voxel-by-voxel heterogeneous spatial distributions of cell size and cell density across whole tumors can still be obtained as shown in Figure S9.

Not only mean cell size $d$ but also the intracellular volume fraction $v_{in}$ can be derived from IMPULSED. The apparent cell density can be calculated from $d$ and $v_{in}$ (28). However, our previous study (34) found that, if the transcytolemmal water exchange cannot be ignored, the intracellular volume fraction is intrinsically underestimated for any biophysical diffusion model that assumes no water exchange (such as IMPULSED and VERDICT). Because cell membrane permeability is likely to increase during treatment-induced apoptosis, the accuracy of cell density derived from IMPULSED may be compromised. Therefore, caution should be expressed in the interpretation of fitted $v_{in}$ values as intracellular volume fraction. Interestingly, $v_{in}$ has been found to correlate well with ground truth cell density (27), which suggests $v_{in}$ might be still a good indicator of cell density although its absolute accuracy is uncertain.

Although only breast cancer was investigated in the current study, IMPULSED can be used to assess other extracranial tumors such as head and neck tumors, wherever a two-compartment model is valid. Note that IMPULSED cannot specifically differentiate cancer cells from other cells. Therefore, the mean cell size obtained using IMPULSED includes all cell types (e.g., cancer cells, stromal cells, and lymphocytes). Our *in vitro* cell studies (Figure 3) suggest IMPULSED has sufficient sensitivity to differentiate small lymphocytes, relatively small cancer cells (leukemia Jarkat), and relatively large breast cancer cells. The relative fraction of these cells will change the mean cell size. If more detailed information on specific types of cells is needed, the two-compartment model may need to be modified. In addition to cell size, other information such as cell shape can be included for such a purpose (49).

To ensure clinical translation, the IMPULSED method presented in the current study was strictly limited to diffusion parameters that are achievable on clinical 3T MRI scanners, such as a maximum gradient strength < 60 mT/m on any single axis and maximum gradient slew rate < 100 mT/m/sec. However, the ability of IMPULSED is not limited by these practical parameters. For example, more advanced MRI hardware such as the Human Connectome gradient coil with a maximum gradient strength 300 mT/m and maximum slew rate 200 mT/m/sec can remarkably improve the ability of diffusion MRI to probe brain microstructure (39). With more advanced hardware, IMPULSED would be capable of acquiring data with higher b values and shorter diffusion times, both of which would enhance the ability of IMPULSED to measure small cell sizes and intracellular diffusion coefficients more reliably.

## Conclusions

A novel and fast IMPULSED imaging method has been successfully developed and validated on clinical MRI scanners for in vivo imaging of mean cell sizes of solid tumors in breast cancer patients. To the best of our knowledge, this is the first clinical study that uses a non-invasive imaging method for spatially mapping distributions of mean cancer cell size of heterogeneous human breast tumors *in vivo*.

## Acknowledgements

The authors thank MR technologists Clair Jones, Leslie McIntosh, Christopher Thompson, and Fuxue Xin for assistance in data acquisitions, Drs. Katy Beckermann and Kirsten Young for collecting lymphocytes. This work was funded by NIH grants K25CA168936, R01CA109106, R01CA173593, UL1TR002243, S10OD021771, U01CA142565, F32CA216942, UL1TR000445, P30 CA068485, and ACS grant IRG#58-009-56.

**References**

1. Kozlowski J, Konarzewski M, Gawelczyk AT. Cell size as a link between noncoding DNA and metabolic rate scaling. P Natl Acad Sci USA 2003;100(24):14080-14085.
2. Baserga R. Is cell size important? Cell cycle 2007;6(7):814-816.
3. Savage VM, Allen AP, Brown JH, Gillooly JF, Herman AB, Woodruff WH, West GB. Scaling of number, size, and metabolic rate of cells with body size in mammals. P Natl Acad Sci USA 2007;104(11):4718-4723.
4. Garcia JH, Yoshida Y, Chen H, Li Y, Zhang ZG, Lian J, Chen S, Chopp M. Progression from ischemic injury to infarct following middle cerebral artery occlusion in the rat. The American Journal of Pathology 1993;142(2):623-635.
5. Liang D, Bhatta S, Gerzanich V, Simard JM. Cytotoxic edema: mechanisms of pathological cell swelling. Neurosurgical focus 2007;22(5):E2-E2.
6. Saraste A, Pulkki K. Morphologic and biochemical hallmarks of apoptosis. Cardiovascular research 2000;45:528-537.
7. Sun L, Sakurai S, Sano T, Hironaka M, Kawashima O, Nakajima T. High-grade neuroendocrine carcinoma of the lung: Comparative clinicopathological study of large cell neuroendocrine carcinoma and small cell lung carcinoma. Pathol Int 2009;59(8):522-529.
8. Brauer M. In vivo monitoring of apoptosis. Prog Neuro-Psychoph 2003;27(2):323-331.
9. Jiang X, McKinley ET, Xie J, Li H, Xu J, Gore JC. In vivo magnetic resonance imaging of treatment-induced apoptosis. Scientific Reports (in revision).
10. Jiang X, Li H, Zhao P, Xie J, Khabele D, Xu J, Gore JC. Early detection of treatment-induced mitotic arrest using temporal diffusion magnetic resonance spectroscopy. Neoplasia 2016;18(6):387-397.
11. Tobkes AI, Nord HJ. Liver-Biopsy - Review of Methodology and Complications. Digest Dis 1995;13(5):267-274.
12. Zhao M, Pipe JG, Bonnett J, Evelhoch JL. Early detection of treatment response by diffusion-weighted 1H-NMR spectroscopy in a murine tumour in vivo. Br J Cancer 1996;73(1):61-64.
13. Sugahara T, Korogi Y, Kochi M, Ikushima I, Shigematu Y, Hirai T, Okuda T, Liang L, Ge Y, Komohara Y, Ushio Y, Takahashi M. Usefulness of diffusion-weighted MRI with echo-planar technique in the evaluation of cellularity in gliomas. J Magn Reson Imaging 1999;9(1):53-60.
14. Gauvain KM, McKinstry RC, Mukherjee P, Perry A, Neil JJ, Kaufman BA, Hayashi RJ. Evaluating pediatric brain tumor cellularity with diffusion-tensor imaging. AJR Am J Roentgenol 2001;177(2):449-454.
15. Ross BD, Moffat BA, Lawrence TS, Mukherji SK, Gebarski SS, Quint DJ, Johnson TD, Junck L, Robertson PL, Muraszko KM, Dong Q, Meyer CR, Bland PH, McConville P, Geng H, Rehemtulla A, Chenevert TL. Evaluation of cancer therapy using diffusion magnetic resonance imaging. Mol Cancer Ther 2003;2(6):581-587.
16. Cory DG, Garroway AN. Measurement of translational displacement probabilities by NMR: an indicator of compartmentation. Magn Reson Med 1990;14(3):435-444.
17. Tanner JE. Transient diffusion in a system partitioned by permeable barriers - application to NMR

measurements with a pulsed field gradient. J Chem Phys 1978;69(4):1748-1754.
18. Szafer A, Zhong J, Gore JC. Theoretical model for water diffusion in tissues. Magn Reson Med 1995;33(5):697-712.
19. van der Toorn A, Syková E, Dijkhuizen RM, Vorísek I, Vargová L, Skobisová E, van Lookeren Campagne M, Reese T, Nicolay K. Dynamic changes in water ADC, energy metabolism, extracellular space volume, and tortuosity in neonatal rat brain during global ischemia. Magnetic resonance in medicine 1996;36:52-60.
20. Padhani AR, Liu G, Koh DM, Chenevert TL, Thoeny HC, Takahara T, Dzik-Jurasz A, Ross BD, Van Cauteren M, Collins D, Hammoud DA, Rustin GJ, Taouli B, Choyke PL. Diffusion-weighted magnetic resonance imaging as a cancer biomarker: consensus and recommendations. Neoplasia 2009;11(2):102-125.
21. Yoshikawa MI, Ohsumi S, Sugata S, Kataoka M, Takashima S, Mochizuki T, Ikura H, Imai Y. Relation between cancer cellularity and apparent diffusion coefficient values using diffusion-weighted magnetic resonance imaging in breast cancer. Radiat Med 2008;26(4):222-226.
22. Squillaci E, Manenti G, Cova M, Di Roma M, Miano R, Palmieri G, Simonetti G. Correlation of diffusion-weighted MR imaging with cellularity of renal tumours. Anticancer Res 2004;24(6):4175-4179.
23. Xu J, Li K, Smith RA, Waterton JC, Zhao P, Chen H, Does MD, Manning HC, Gore JC. Characterizing tumor response to chemotherapy at various length scales using temporal diffusion spectroscopy. PLoS One 2012;7(7):e41714.
24. Panagiotaki E, Walker-Samuel S, Siow B, Johnson SP, Rajkumar V, Pedley RB, Lythgoe MF, Alexander DC. Noninvasive quantification of solid tumor microstructure using VERDICT MRI. Cancer Res 2014;74(7):1902-1912.
25. Panagiotaki E, Chan RW, Dikaios N, Ahmed HU, O'Callaghan J, Freeman A, Atkinson D, Punwani S, Hawkes DJ, Alexander DC. Microstructural characterization of normal and malignant human prostate tissue with vascular, extracellular, and restricted diffusion for cytometry in tumours magnetic resonance imaging. Investigative radiology 2015;50(4):218-227.
26. Bonet-Carne E, Johnston E, Daducci A, Jacobs JG, Freeman A, Atkinson D, Hawkes DJ, Punwani S, Alexander DC, Panagiotaki E. VERDICT-AMICO: Ultrafast fitting algorithm for non-invasive prostate microstructure characterization. NMR in biomedicine 2019;32(1):e4019.
27. Jiang X, Li H, Xie J, Zhao P, Gore JC, Xu J. Quantification of cell size using temporal diffusion spectroscopy. Magn Reson Med 2016;75(3):1076-1085.
28. Jiang X, Li H, Xie J, McKinley ET, Zhao P, Gore JC, Xu J. In vivo imaging of cancer cell size and cellularity using temporal diffusion spectroscopy. Magn Reson Med 2017;78(1):156-164.
29. Reynaud O, Winters KV, Hoang DM, Wadghiri YZ, Novikov DS, Kim SG. Pulsed and oscillating gradient MRI for assessment of cell size and extracellular space (POMACE) in mouse gliomas. NMR in biomedicine 2016;29(10):1350-1363.
30. Van AT, Holdsworth SJ, Bammer R. In vivo investigation of restricted diffusion in the human brain with optimized oscillating diffusion gradient encoding. Magn Reson Med 2014;71(1):83-94.
31. Baron CA, Beaulieu C. Oscillating gradient spin-echo (OGSE) diffusion tensor imaging of the human brain. Magn Reson Med 2014;72(3):726-736.

32. Gore JC, Xu J, Colvin DC, Yankeelov TE, Parsons EC, Does MD. Characterization of tissue structure at varying length scales using temporal diffusion spectroscopy. NMR Biomed 2010;23(7):745-756.
33. Xu J, Does MD, Gore JC. Quantitative characterization of tissue microstructure with temporal diffusion spectroscopy. J Magn Reson 2009;200(2):189-197.
34. Li H, Jiang X, Xie J, Gore JC, Xu J. Impact of transcytolemmal water exchange on estimates of tissue microstructural properties derived from diffusion MRI. Magn Reson Med 2017;77(6):2239-2249.
35. Aggarwal M, Jones MV, Calabresi PA, Mori S, Zhang J. Probing mouse brain microstructure using oscillating gradient diffusion MRI. Magn Reson Med 2012;67(1):98-109.
36. Bailey C, Siow B, Panagiotaki E, Hipwell JH, Mertzanidou T, Owen J, Gazinska P, Pinder SE, Alexander DC, Hawkes DJ. Microstructural models for diffusion MRI in breast cancer and surrounding stroma: an ex vivo study. NMR in biomedicine 2017;30(2):e3679.
37. Partridge SC, Ziadloo A, Murthy R, White SW, Peacock S, Eby PR, DeMartini WB, Lehman CD. Diffusion tensor MRI: Preliminary anisotropy measures and mapping of breast tumors. Journal of Magnetic Resonance Imaging 2010;31:339-347.
38. Xu J, Li H, Harkins KD, Jiang X, Xie J, Kang H, Does MD, Gore JC. Mapping mean axon diameter and axonal volume fraction by MRI using temporal diffusion spectroscopy. Neuroimage 2014;103:10-19.
39. Setsompop K, Kimmlingen R, Eberlein E, Witzel T, Cohen-Adad J, McNab JA, Keil B, Tisdall MD, Hoecht P, Dietz P, Cauley SF, Tountcheva V, Matschl V, Lenz VH, Heberlein K, Potthast A, Thein H, Van Horn J, Toga A, Schmitt F, Lehne D, Rosen BR, Wedeen V, Wald LL. Pushing the limits of in vivo diffusion MRI for the Human Connectome Project. NeuroImage 2013;80:220-233.
40. Xu J, Does MD, Gore JC. Numerical study of water diffusion in biological tissues using an improved finite difference method. Phys Med Biol 2007;52(7):N111-126.
41. Xu J, Does MD, Gore JC. Sensitivity of MR diffusion measurements to variations in intracellular structure: effects of nuclear size. Magn Reson Med 2009;61(4):828-833.
42. Boyle W, Chow A. Isolation of human lymphocytes by a Ficoll barrier method. Transfusion 1969;9(3):151-155.
43. Ali R, Gooding M, Szilágyi T, Vojnovic B, Christlieb M, Brady M. Automatic segmentation of adherent biological cell boundaries and nuclei from brightfield microscopy images. Machine Vision and Applications 2012;23(4):607-621.
44. Neeman M, Freyer JP, Sillerud LO. A simple method for obtaining cross-term-free images for diffusion anisotropy studies in NMR microimaging. Magnetic Resonance in Medicine 1991;21:138-143.
45. Jenkinson M, Smith S. A global optimisation method for robust affine registration of brain images. Medical image analysis 2001;5(2):143-156.
46. Jenkinson M, Beckmann CF, Behrens TE, Woolrich MW, Smith SM. Fsl. Neuroimage 2012;62(2):782-790.
47. Del Monte U. Does the cell number 10(9) still really fit one gram of tumor tissue? Cell cycle 2009;8(3):505-506.

48. Assaf Y, Blumenfeld-Katzir T, Yovel Y, Basser PJ. AxCaliber: a method for measuring axon diameter distribution from diffusion MRI. Magn Reson Med 2008;59(6):1347-1354.
49. Szczepankiewicz F, van Westen D, Englund E, Westin CF, Stahlberg F, Latt J, Sundgren PC, Nilsson M. The link between diffusion MRI and tumor heterogeneity: Mapping cell eccentricity and density by diffusional variance decomposition (DIVIDE). Neuroimage 2016.

# Supplemental Materials

# for

# "Magnetic resonance imaging of mean cell size in human breast tumors"

Junzhong Xu, Xiaoyu Jiang, Hua Li, Lori R. Arlinghaus, Eliot T. McKinley, Sean P Devan, Benjamin M Hardy, Jingping Xie, Hakmook Kang, Anuradha B. Chakravarthy, John C. Gore

**Analytical expressions of intracellular diffusion signal**

Diffusion-weighted MRI (DWI) signal attenuation assuming a Gaussian phase distribution can be described as (1)

$$S = S_0 \exp(-\phi) \, , \qquad [S1]$$

where $S_0$ is the T2-weighted non-diffusion-weighted signal, and signal echo attenuation $\phi$ can be written as

$$\phi = \frac{\gamma^2}{2} \sum_k B_k \int_0^{TE} dt_1 \int_0^{TE} dt_2 \exp(-a_k D |t_2 - t_1|) g(t_1) g(t_2) \qquad [S2]$$

where $\gamma$ is the gyromagnetic ratio, $TE$ = echo time, $g(t)$ = effective time-varying diffusion gradient, $D$ = intrinsic intracellular diffusion coefficient, $B_k$ and $a_k$ are microstructure dependent coefficients which have been derived previously for simple geometries such as cylinders and spheres (1). The importance of Eq.[S2] is the separation of microstructural ($B_k$ and $a_k$) and experimental parameters ($g(t)$), so that the analytical equation of DWI signals of diffusion sequences with any gradient waveform can be derived, such as the sine and cosine-modulated (2) and sine-modulated trapezoidal (3) gradient waveforms. For the present work we derived analytical expressions of intracellular DWI signals using the diffusion gradient waveforms ($g(t)$) that are the same as used on clinical MRI scanners, including the finite duration of gradient rise time (i.e., $t_r$).

For practical PGSE sequences with trapezoid-shaped gradient waveforms, the analytical expression of the intracellular diffusion signal attenuation is given as (3)

$$\phi = \gamma^2 G^2 \sum_k \frac{B_k}{a_k^4 D^4 t_r^2} \begin{Bmatrix} 2\exp(-a_k D t_p) - 4\exp(-a_k D t_r) - 4\exp(-\Delta a_k D) - 4 a_k D t_r + 2\exp[-a_k D(\Delta - t_r)] \\ +2\exp[-a_k D(\Delta + t_r)] - \exp[-a_k D(\Delta - t_p)] - \exp[-a_k D(\Delta + t_p)] - 4\exp[-a_k D(t_r + t_p)] \\ +2\exp[-a_k D(2 t_r + t_p)] + (4/3) a_k^3 D^3 t_r^3 + 2 a_k^3 D^3 t_r^2 t_p + 2\exp[-a_k D(\Delta - t_r - t_p)] \\ +2\exp[-a_k D(\Delta + t_r + t_p)] - \exp[-a_k D(\Delta - 2 t_r - t_p)] - \exp[-a_k D(\Delta + 2 t_r + t_p)] + 4 \end{Bmatrix} \qquad [S3]$$

where $t_r$ is the gradient rise time and $t_p$ is the duration of each gradient plateau. Note that the analytical expression of the attenuation of the PGSE DWI signal was derived previously based on the square waveform for simplicity (i.e., infinitely fast gradient slew rate) as (1)

$$\phi = 2 \left( \frac{\gamma g}{D} \right)^2 \sum_k \frac{B_k}{a_k^2} \left\{ a_k D \delta - 1 + e^{-a_k D \delta} + e^{-a_k D \Delta} (1 - \cosh(a_k D \delta)) \right\} \, . \qquad [S4]$$

Because the gradient slew rate is usually very limited on clinical MRI scanners, the neglect of the finite duration of the gradient ramp decreases the accuracy of estimates of cell size. To validate Eqs. [S3], we used computer simulations which incorporate the true trapezoidal waveform used for clinical MRI scans. DWI signals inside a perfectly impermeable sphere were simulated for multiple cell diameters. Other simulation parameters were the same as those shown in Table 2 except that b = 1000 s/mm². Note that we have previously shown that our computer simulation is capable of providing sufficient accuracy (<1% in ADC) to predict DWI signals (4). Figure S1 shows a comparison of simulated intracellular PGSE signals vs the analytical forms with a trapezoidal waveform (Eq.[S3]) and a square waveform (Eq.[S4]). Eq. [S3] yields an error < 4% across a large range of cell sizes from 3 to 20 µm. By contrast, Eq. [S4] yields errors > 10% for any cell size larger than 10 µm, which includes the range of typical cancer cell sizes. Therefore, it is important to use the analytical expressions derived using the true gradient waveforms (i.e., with a finite gradient slew rate) in order to achieve accurate estimates of cell size.

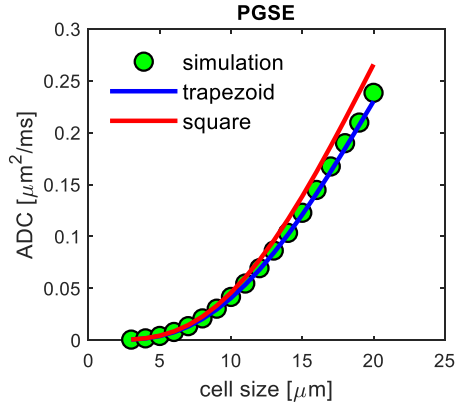

Figure S1 Comparison of intracellular PGSE DWI signals obtained from simulation vs analytical forms with trapezoid waveform (Eq.[S3]) an square waveform (Eq.[S4])

A previous study (3) has reported analytical equations for sine-modulated trapezoidal oscillating gradients. However, the cosine-modulated trapezoidal waveform provides much shorter effective diffusion times and hence were used in the current study. We obtained the analytical expressions of the DWI signal attenuation for the cosine-modulated trapezoidal OGSE sequence shown in Figure 1 as

$$\phi = \gamma^2 G^2 \sum_k \frac{B_k}{a_k^4 D^4 t_r^2} \begin{Bmatrix} 4\exp(-a_k D t_p) - 4\exp(-a_k D t_r) - 4\exp(-2a_k D t_r) - 8\exp(-a_k D \Delta) - 12 a_k D t_r + 2\exp[-a_k D(\Delta - t_r)] \\ +2\exp[-a_k D(\Delta + t_r)] + 2\exp[-a_k D(\Delta + 2t_r)] + 2\exp[-a_k D(\Delta - 2t_r)] - 2\exp[-a_k D(\Delta - t_p)] \\ -2\exp[-a_k D(\Delta + t_p)] - 4\exp[-a_k D(t_r + t_p)] + 2\exp[-a_k D(t_r + 2t_p)] - 4\exp[-a_k D(2t_r + t_p)] \\ +4\exp[-a_k D(3t_r + t_p)] - 4\exp[-a_k D(3t_r + 2t_p)] - 4\exp[-a_k D(3t_r + 3t_p)] + 4\exp[-a_k D(4t_r + 3t_p)] \\ +2\exp[-a_k D(5t_r + 2t_p)] + 4\exp[-a_k D(5t_r + 3t_p)] + 2\exp[-a_k D(5t_r + 4t_p)] - 4\exp[-a_k D(6t_r + 3t_p)] \\ -4\exp[-a_k D(6t_r + 4t_p)] + 2\exp[-a_k D(7t_r + 4t_p)] + 6a_k^3 D^3 t_r^3 + 8a_k^3 D^3 t_r^2 t_p + 2\exp[-a_k D(\Delta - t_r - t_p)] \\ -\exp[-a_k D(\Delta - t_r - 2t_p)] + 2\exp[-a_k D(\Delta + t_r + t_p)] + 2\exp[-a_k D(\Delta - 2t_r - t_p)] \\ -\exp[-a_k D(\Delta + t_r + 2t_p)] + 2\exp[-a_k D(\Delta + 2t_r + t_p)] - 2\exp[-a_k D(\Delta - 3t_r - t_p)] \\ -2\exp[-a_k D(\Delta + 3t_r + t_p)] + 2\exp[-a_k D(\Delta + 3t_r + 2t_p)] + 2\exp[-a_k D(\Delta - 3t_r - 2t_p)] \\ +2\exp[-a_k D(\Delta + 3t_r + 3t_p)] + 2\exp[-a_k D(\Delta - 3t_r - 3t_p)] - 2\exp[-a_k D(\Delta + 4t_r + 3t_p)] \\ -2\exp[-a_k D(\Delta - 4t_r - 3t_p)] - \exp[-a_k D(\Delta + 5t_r + 2t_p)] - \exp[-a_k D(\Delta - 5t_r - 2t_p)] \\ -2\exp[-a_k D(\Delta + 5t_r + 3t_p)] - 2\exp[-a_k D(\Delta - 5t_r - 3t_p)] - \exp[-a_k D(\Delta + 5t_r + 4t_p)] \\ -\exp[-a_k D(\Delta - 5t_r - 4t_p)] + 2\exp[-a_k D(\Delta + 6t_r + 3t_p)] + 2\exp[-a_k D(\Delta - 6t_r - 3t_p)] \\ +2\exp[-a_k D(\Delta + 6t_r + 4t_p)] + 2\exp[-a_k D(\Delta - 6t_r - 4t_p)] - \exp[-a_k D(\Delta + 7t_r + 4t_p)] \\ -\exp[-a_k D(\Delta - 7t_r - 4t_p)] + 8 \end{Bmatrix} \quad [S5]$$

Recall that $t_p$ is the duration of the first gradient plateau in the gradient waveform.

For the trapezoidal OGSE sequence with N=2, the analytical expression of the DWI signal attenuation is

$$\phi = \gamma^2 G^2 \sum_k \frac{B_k N}{a_k^4 D^4 t_r^2} \left\{ \begin{array}{l} 4\exp(-a_k D t_p) - 4\exp(-a_k D t_r) - 8\exp(-2a_k D t_r) - 12\exp(-a_k D \Delta) - 20 a_k D t_r + 2\exp[-a_k D(\Delta - t_r)] \\ +2\exp[-a_k D(\Delta + t_r)] + 4\exp[-a_k D(\Delta + 2t_r)] + 4\exp[-a_k D(\Delta - 2t_r)] - 2\exp[-a_k D(\Delta - t_p)] \\ -2\exp[-a_k D(\Delta + t_p)] - 4\exp[-a_k D(t_r + t_p)] + 6\exp[-a_k D(t_r + 2t_p)] - 4\exp[-a_k D(2t_r + t_p)] \\ +4\exp[-a_k D(3t_r + t_p)] - 12\exp[-a_k D(3t_r + 2t_p)] - 4\exp[-a_k D(3t_r + 3t_p)] + 4\exp[-a_k D(4t_r + 3t_p)] \\ +6\exp[-a_k D(5t_r + 2t_p)] - 4\exp[-a_k D(4t_r + 4t_p)] + 4\exp[-a_k D(5t_r + 3t_p)] - 4\exp[-a_k D(6t_r + 3t_p)] \\ +8\exp[-a_k D(6t_r + 4t_p)] + 4\exp[-a_k D(6t_r + 5t_p)] - 4\exp[-a_k D(7t_r + 5t_p)] - 4\exp[-a_k D(8t_r + 4t_p)] \\ +2\exp[-a_k D(7t_r + 6t_p)] - 4\exp[-a_k D(8t_r + 5t_p)] + 4\exp[-a_k D(9t_r + 5t_p)] - 4\exp[-a_k D(9t_r + 6t_p)] \\ -4\exp[-a_k D(9t_r + 7t_p)] + 4\exp[-a_k D(10t_r + 7t_p)] + 2\exp[-a_k D(11t_r + 6t_p)] + 4\exp[-a_k D(11t_r + 7t_p)] \\ +2\exp[-a_k D(11t_r + 8t_p)] - 4\exp[-a_k D(12t_r + 7t_p)] - 4\exp[-a_k D(12t_r + 8t_p)] + 2\exp[-a_k D(13t_r + 8t_p)] \\ +(38/3)a_k^3 D^3 t_r^3 + 16 a_k^3 D^3 t_r^2 t_p + 2\exp[-a_k D(\Delta - t_r - t_p)] - 3\exp[-a_k D(\Delta - t_r - 2t_p)] \\ +2\exp[-a_k D(\Delta + t_r + t_p)] + 2\exp[-a_k D(\Delta - 2t_r - t_p)] - 3\exp[-a_k D(\Delta + t_r + 2t_p)] \\ +2\exp[-a_k D(\Delta + 2t_r + t_p)] - 2\exp[-a_k D(\Delta - 3t_r - t_p)] - 2\exp[-a_k D(\Delta + 3t_r + t_p)] \\ +6\exp[-a_k D(\Delta + 3t_r + 2t_p)] + 6\exp[-a_k D(\Delta - 3t_r - 2t_p)] + 2\exp[-a_k D(\Delta + 3t_r + 3t_p)] \\ +2\exp[-a_k D(\Delta - 3t_r - 3t_p)] - 2\exp[-a_k D(\Delta + 4t_r + 3t_p)] - 2\exp[-a_k D(\Delta - 4t_r - 3t_p)] \\ -3\exp[-a_k D(\Delta + 5t_r + 2t_p)] - 3\exp[-a_k D(\Delta - 5t_r - 2t_p)] + 2\exp[-a_k D(\Delta + 4t_r + 4t_p)] \\ +2\exp[-a_k D(\Delta - 4t_r - 4t_p)] - 2\exp[-a_k D(\Delta + 5t_r + 3t_p)] - 2\exp[-a_k D(\Delta - 5t_r - 3t_p)] \\ +2\exp[-a_k D(\Delta + 6t_r + 3t_p)] + 2\exp[-a_k D(\Delta - 6t_r - 3t_p)] - 4\exp[-a_k D(\Delta + 6t_r + 4t_p)] \\ -4\exp[-a_k D(\Delta - 6t_r - 4t_p)] - 2\exp[-a_k D(\Delta + 6t_r + 5t_p)] - 2\exp[-a_k D(\Delta - 6t_r - 5t_p)] \\ +2\exp[-a_k D(\Delta + 7t_r + 5t_p)] + 2\exp[-a_k D(\Delta - 7t_r - 5t_p)] + 2\exp[-a_k D(\Delta + 8t_r + 4t_p)] \\ +2\exp[-a_k D(\Delta - 8t_r - 4t_p)] - \exp[-a_k D(\Delta + 7t_r + 6t_p)] - \exp[-a_k D(\Delta - 7t_r - 6t_p)] \\ +2\exp[-a_k D(\Delta + 8t_r + 5t_p)] + 2\exp[-a_k D(\Delta - 8t_r - 5t_p)] - 2\exp[-a_k D(\Delta + 9t_r + 5t_p)] \\ -2\exp[-a_k D(\Delta - 9t_r - 5t_p)] + 2\exp[-a_k D(\Delta + 9t_r + 6t_p)] + 2\exp[-a_k D(\Delta - 9t_r - 6t_p)] \\ +2\exp[-a_k D(\Delta + 9t_r + 7t_p)] + 2\exp[-a_k D(\Delta - 9t_r - 7t_p)] - 2\exp[-a_k D(\Delta + 10t_r + 7t_p)] \\ -2\exp[-a_k D(\Delta - 10t_r - 7t_p)] - \exp[-a_k D(\Delta + 11t_r + 6t_p)] - \exp[-a_k D(\Delta - 11t_r - 6t_p)] \\ -2\exp[-a_k D(\Delta + 11t_r + 7t_p)] - 2\exp[-a_k D(\Delta - 11t_r - 7t_p)] - \exp[-a_k D(\Delta + 11t_r + 8t_p)] \\ -\exp[-a_k D(\Delta - 11t_r - 8t_p)] + 2\exp[-a_k D(\Delta + 12t_r + 7t_p)] + 2\exp[-a_k D(\Delta - 12t_r - 7t_p)] \\ +2\exp[-a_k D(\Delta + 12t_r + 8t_p)] + 2\exp[-a_k D(\Delta - 12t_r - 8t_p)] - \exp[-a_k D(\Delta + 13t_r + 8t_p)] \\ -\exp[-a_k D(\Delta - 13t_r - 8t_p)] + 12 \end{array} \right\} \quad [S6]$$

Figure S2 shows a comparison of intracellular DWI signals obtained from simulation vs analytical expressions using trapezoidal OGSE sequences with N=1 (Eq.[S5]) and N=2 (Eq.[S6]). All errors are smaller than ~7% in ADC.

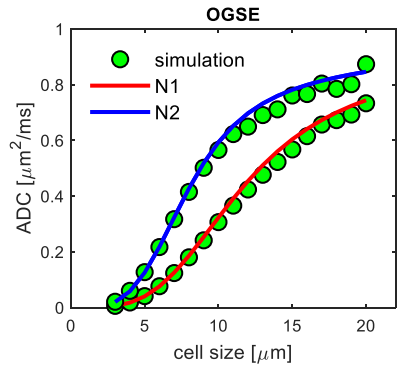

Figure S2 Comparison of intracellular DWI signals obtained from simulation vs analytical expressions with trapezoidal OGSE sequences and N=1 (Eq.[S5]) and N=2 (Eq.[S6])

**Computer simulations with SNR = 50**

Both the accuracy and precision of IMPULSED fitted parameters are significantly improved if SNR = 50 as shown in Figure S3. Particularly, the coefficient of variances of both $d$ and $v_{in}$ decreases to 10% when the ground truth $d$ > 8 μm, indicating both parameters may be fit reliably if SNR is sufficient for the IMPULSED method on clinical human scanners.

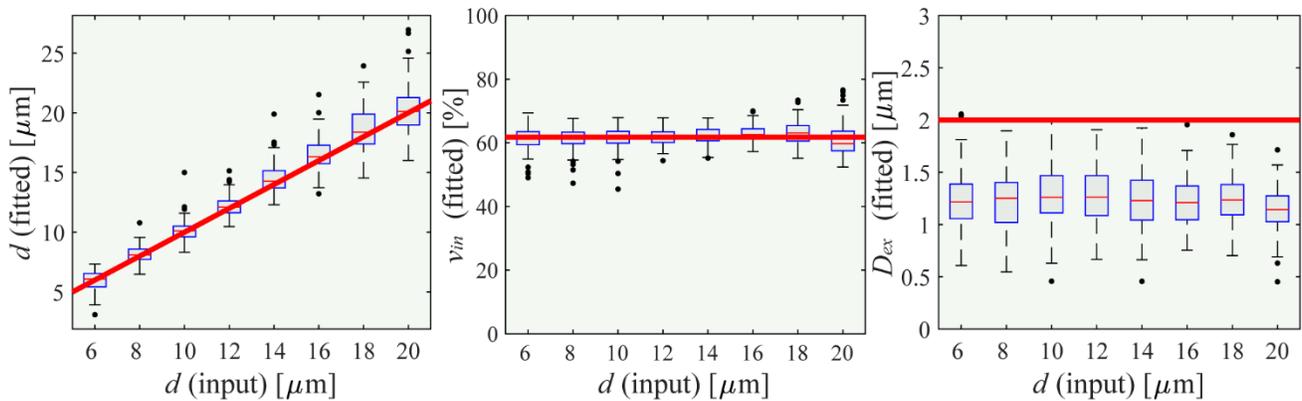

Figure S3 Simulated influence of noise with SNR = 50 on IMPULSED derived metrics. For each real input $d$, the fittings were repeated 100 times each with different noises but with the same SNR level. The red solid lines represent the ground truth values inputted in the simulations, boxes represent ranges between the 25th and 75th percentiles, and dots are outliers.

**Influence of $D_{in}$**

The intracellular diffusivity $D_{in}$ has been considered as a free fitting parameter in previous pre-clinical studies (5-8). Although the diffusion times $t_{diff}$ (10 and 5 ms) used in the IMPULSED acquisitions are shorter than those in conventional PGSE acquisition (> 30 ms), they are much longer than those (e.g., 1 ms) used in the pre-clinical studies *in vivo*. Therefore, the sensitivity to the intracellular diffusivity is reduced and then it is questionable how accurate and reliable $D_{in}$ can be fit on clinical human MRI scanners. Figure S4 shows the fitting results when $D_{in}$ is considered as a free fitting parameter. $D_{in}$ cannot be reliably fit from IMPULSED. Presumably, this is due to the limitation on the shortest achievable diffusion time on clinical MRI scanners. Since the sensitivity of IMPULSED on clinical MRI scanners is significantly reduced to $D_{in}$, it is possible to consider $D_{in}$ as a fixed constant value in all IMPULSED fitting. This reduces the number of free fitting parameters, which in turn stabilizes the fitting.

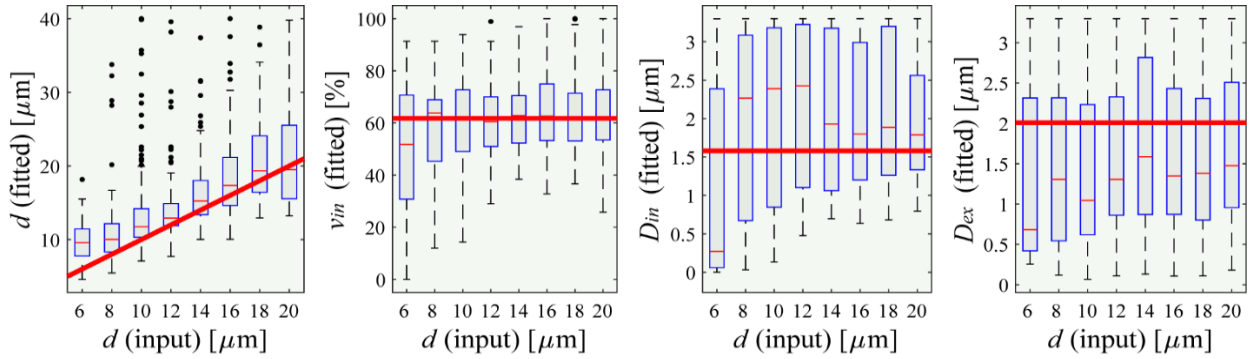

Figure S4 Simulated IMPULSED derived metrics with SNR = 20 if $D_{in}$ is a free fitting parameter.

However, $D_{in}$ is usually unknown in practice in vivo and may change during apoptosis or mitotic arrest (6). It is unclear if different choices of fixed $D_{in}$ values in the data analysis may influence the final fitting results of the IMPULSED method. We therefore performed computer simulations using a single $D_{in}$ value of 1.58 µm²/ms (6) to synthesize diffusion MRI signals, but used a wide range of different $D_{in}$ values (1.0 – 2.2 µm²/ms) in the data analyses to investigate how the choices of fixed $D_{in}$ in data analysis may affect the fitting results. Figure S5 shows the simulated results of fitted IMPULSED parameters with different $D_{in}$ values. The fitted $d$ and $D_{in}$ are slightly biased and the corresponding $D_{ex}$ has much larger uncertainties only when the ground truth mean cell size is > 14 µm and fixed $D_{in}$ < 1.4 µm²/ms. This is expected because a large cell size and slow $D_{in}$ will lead to a strong diffusion time dependence in the intracellular space and a fixed $D_{in}$ will obviously bias fittings. However, fittings of all other combinations of $d$ and fixed $D_{in}$ are not significantly affected by the choices of $D_{in}$ used in the data analysis. This confirms again that the IMPULSED fitting with the acquisition protocol proposed in the current work is insensitive to $D_{in}$ so $D_{in}$ can be chosen as a fixed constant value in the data analysis. Since $D_{in}$ is usually unknown, it is preferable to choose a relatively larger $D_{in}$ in the fittings compared with the ground truth value and this still provide robust fittings of microstructural parameters as shown in Figure S5.

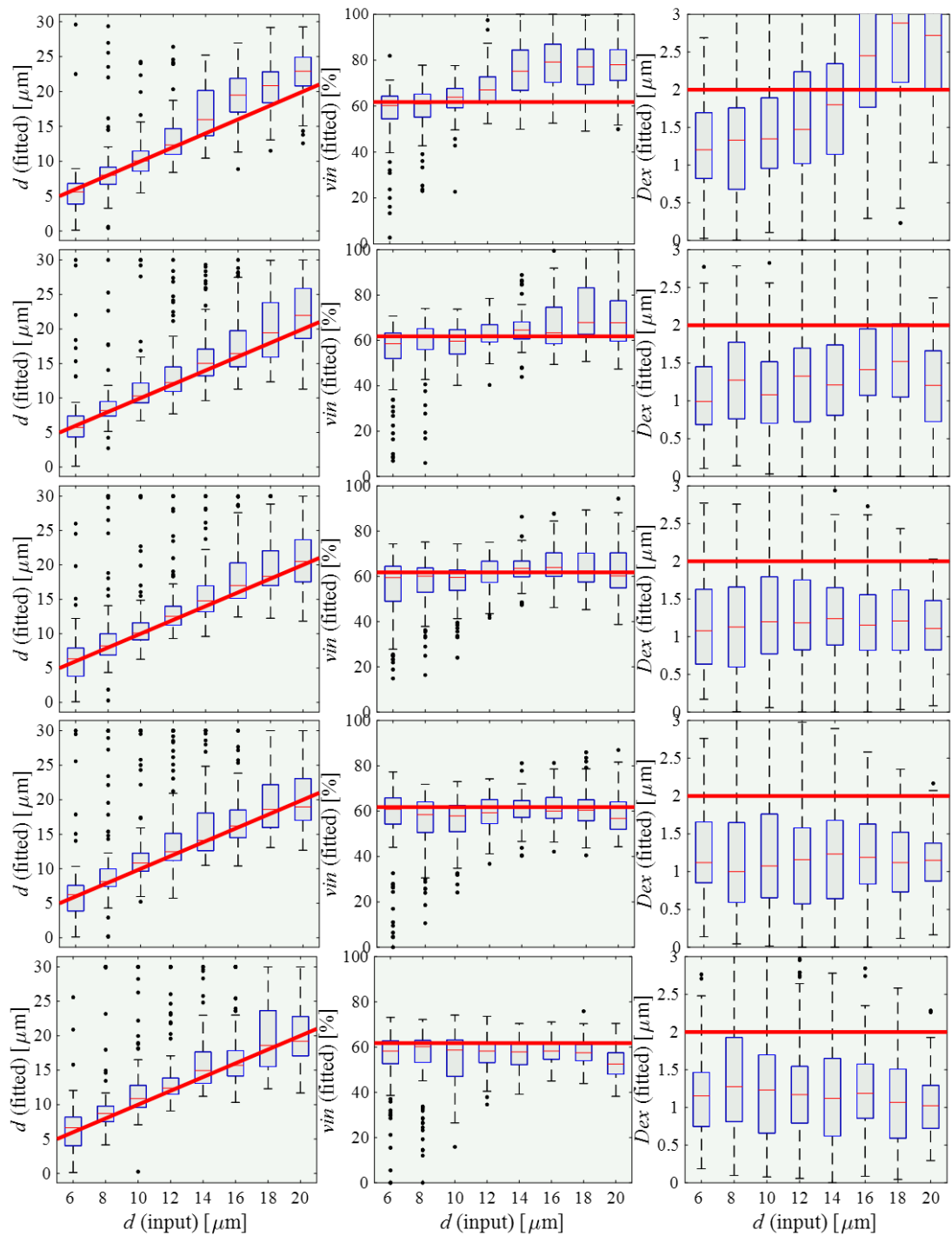

Figure S5 Simulated influence of noise with SNR = 20 on IMPULSED derived metrics with different choices of fixed $D_{in}$ values in the data analyses. From the top to bottom, fixed $D_{in}$ = 1.0, 1.4, 1.58 (ground truth), 1.8, 2.2 μm²/ms.

**Cell segmentation of light microscopy images**

Each field of view was imaged at 40X magnification with the focus set slightly above, below, and equal to the optimal focal plane, and the combination of the images provide better identification of cell boundaries (9). The detailed pipeline is shown below, and a representative FOV is shown in Figure S6:

1. Reading the images;
2. Calculating the difference between two out-of-focus images for enhanced contrast of cell boundary;
3. Calculating the difference between in-focus and out-of-focus images for background elimination;
4. Identifying the entire cell region by thresholding and morphological transforming images calculated from step 2;
5. Identifying seeds for watershedding from images calculated from step 3;
6. Splitting clumping cells using watershedding;
7. Correcting over-segmented cells manually;
8. Calculating cell area and converting to cell diameter for each cell.

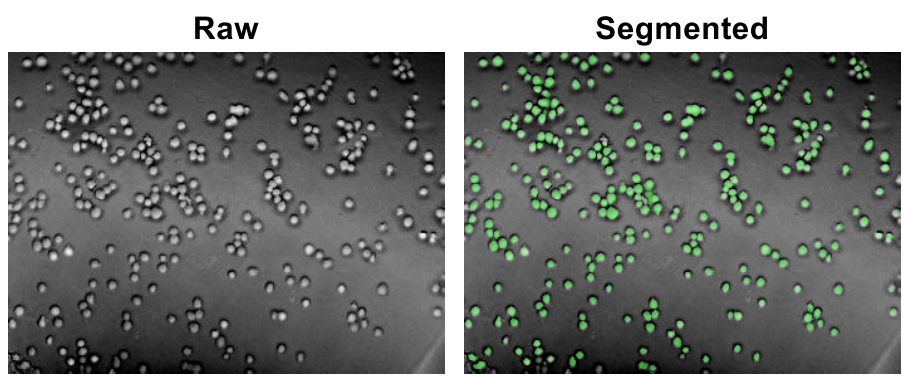

Figure S6 A representative FOV of light microscopy of cells in vitro. The left is the raw image and the right shows the segmented images with identified cells highlighted in green.

**Visualization of cell boundaries in histology**

The cell membrane and nucleus stained images were analyzed using CellProfiler™ (http://www.cellprofiler.org/) to obtain quantitative information on cell sizes. The complete processing pipeline includes: illumination correction, foreground objects identification, and splitting of clumped cells by watershedding. Finally, the cell segmentation results obtained by CellProfiler were manually corrected for over-segmented cell bodies. The corresponding CellProfiler pipeline file is available upon request. A representative image is shown in Figure S7.

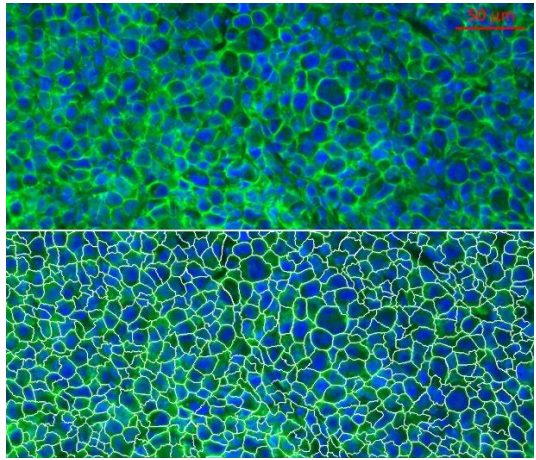

Figure S7 A representative image of an MDA-MB-231 breast tumor in mouse. The top is the raw image, and the bottom is the images after the processing of CellProfiler.

**Diffusion time dependence in human tumors in vivo**

Figure S8 shows the dependence of ADC of a human breast tumor on diffusion times. Except that b = 300 s/mm$^2$ for $t_{diff}$ = 5 ms, all other ADCs were obtained using b = 1000 s/mm$^2$. It is evident that ADC increases with decreasing diffusion time under 10 ms, but ADC does not vary significantly in the $t_{diff}$ range of 30 to 70 ms. This suggests that PGSE measurements with long diffusion times are less sensitive to structural differences at cellular scale, while OGSE measurements with shorter diffusion times are sensitive to the specific sizes of restricting distances (i.e., cell size). Therefore, the IMPULSED method incorporates PGSE measurements with $t_{diff}$ = 70 ms only without the need of PGSE experiments of $t_{diff}$ = 30 or 50 ms. This in turn reduces the acquisition time.

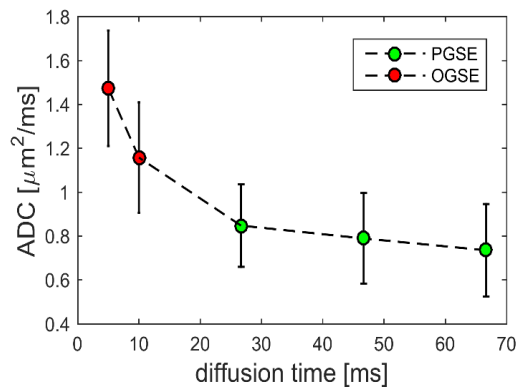

Figure S8 Tumor ADC dependence on diffusion time. The error-bars represent the standard deviations over the tumor region shown in Fig 4.

**Breast cancer patients**

Information on breast cancer patients and breast tumors are tabulated below.

Table S1 Information on breast cancer patients and their tumor types.

| Patient | age | Tumor grade | Clinical stage | ER | PR | HER2-Neu |
|---|---|---|---|---|---|---|
| 1 | 61 | Low | IIA | + | + | - |
| 2 | 58 | Intermediate | IIA | + | + | - |
| 3 | 52 | High | IIB | - | - | + |
| 4 | 68 | low | I | + | + | - |
| 5 | 49 | Intermediate | I | + | + | equivocal |
| 6 | 55 | low | IIA | + | + | - |
| 7 | 44 | High | IIA | + | + | + |

Figure S9 summarizes the histograms of all fitted IMPULSED metrics for seven patients. Although different patients and breast tumors show different histograms of the IMPULSED metrics, the peaks of $d$ are all in the range of 12 – 18 μm and $v_{in}$ are in the range of 25 – 40 %. Values of $v_{in}$ may be underestimated due to transcytolemmal water change, but this does not affect $d$ (8). $D_{ex}$ show broader ranges of distribution compared with those of $d$ and $v_{in}$. This may be due to their larger fitting variations as predicted by the simulation results (Figure 1).

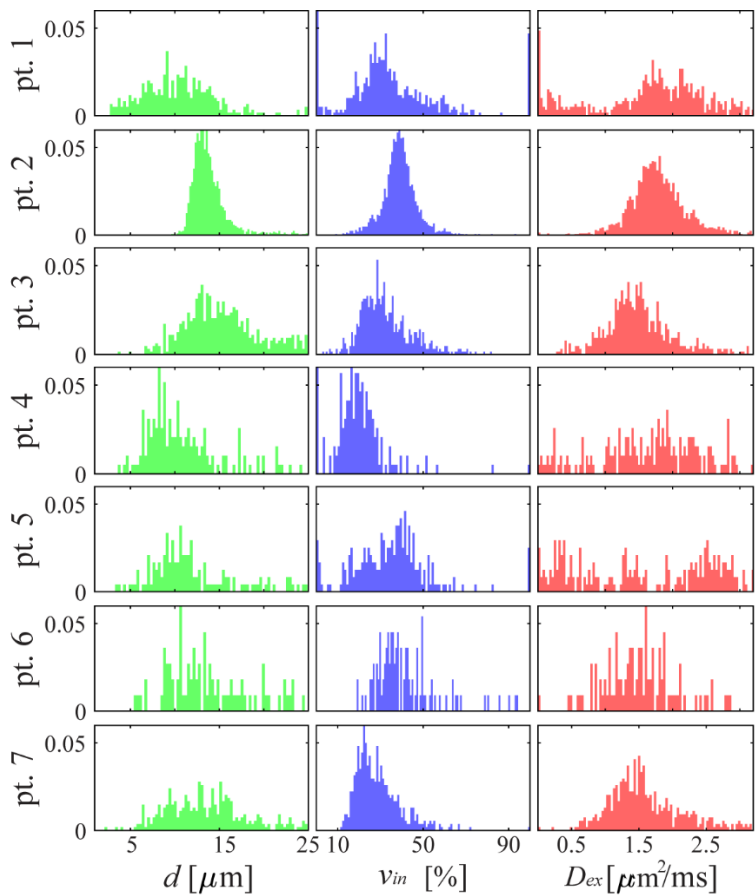

Figure S9 A summary of histograms of all fitted IMPULSED metrics (columns) of all seven patients (rows).

**References**

1. Stepisnik J. Time-dependent self-diffusion by NMR spin-echo. Physica B 1993;183:343-50
2. Xu J, Does MD, Gore JC. Quantitative characterization of tissue microstructure with temporal diffusion spectroscopy. J Magn Reson 2009;200:189-97
3. Ianus A, Siow B, Drobnjak I, Zhang H, Alexander DC. Gaussian phase distribution approximations for oscillating gradient spin echo diffusion MRI. J Magn Reson 2013;227:25-34
4. Xu J, Does MD, Gore JC. Numerical study of water diffusion in biological tissues using an improved finite difference method. Phys Med Biol 2007;52:N111-26
5. Jiang X, Li H, Xie J, Zhao P, Gore JC, Xu J. Quantification of cell size using temporal diffusion spectroscopy. Magn Reson Med 2016;75:1076-85
6. Jiang X, Li H, Zhao P, Xie J, Khabele D, Xu J, *et al.* Early detection of treatment-induced mitotic arrest using temporal diffusion magnetic resonance spectroscopy. Neoplasia 2016;18:387-97
7. Jiang X, Li H, Xie J, McKinley ET, Zhao P, Gore JC, *et al.* In vivo imaging of cancer cell size and cellularity using temporal diffusion spectroscopy. Magn Reson Med 2017;78:156-64
8. Li H, Jiang X, Xie J, Gore JC, Xu J. Impact of transcytolemmal water exchange on estimates of tissue microstructural properties derived from diffusion MRI. Magn Reson Med 2017;77:2239-49
9. Ali R, Gooding M, Szilágyi T, Vojnovic B, Christlieb M, Brady M. Automatic segmentation of adherent biological cell boundaries and nuclei from brightfield microscopy images. Machine Vision and Applications 2012;23:607-21